%% file: paper.tex
\documentclass[sigconf]{acmart}

\AtBeginDocument{%
  \providecommand\BibTeX{{%
    \normalfont B\kern-0.5em{\scshape i\kern-0.25em b}\kern-0.8em\TeX}}}

\setcopyright{acmcopyright}
\copyrightyear{2023}
\acmYear{2023}
\acmDOI{XXXXXXX.XXXXXXX}
\usepackage{xspace}
\usepackage[subtle]{savetrees} 

\definecolor{ballblue}{rgb}{0.13, 0.67, 0.8}
\definecolor{bittersweet}{rgb}{1.0, 0.44, 0.37}

\newcommand{\update}[1]{{\color{blue}#1}}
\newcommand{\vmd}{\textsc{VM dataset}\xspace}
\newcommand{\disd}{\textsc{DISTRIB dataset}\xspace}

\newcommand{\capd}{\textsc{CAP dataset}\xspace}

\newcommand{\test}{\textsc{VM test dataset}\xspace}
\newcommand{\testn}{\textsc{VM-9 test dataset}\xspace}
\newcommand{\testf}{\textsc{VM-14 test dataset}\xspace}
\newcommand{\testgs}{\textsc{GL test dataset}\xspace}
\usepackage{wrapfig}
\usepackage{subcaption}
\usepackage{color, colortbl}
\definecolor{gris1}{gray}{0.85}
\usepackage{xcolor}
\definecolor{color1}{RGB}{63,87,101}
\definecolor{color2}{RGB}{213,232,212}
\definecolor{Gray}{gray}{0.9}
\definecolor{LightCyan}{rgb}{0.88,1,1}
\usepackage{url}
\copyrightyear{2023} 
\acmYear{2023} 
\setcopyright{acmlicensed}\acmConference[WWW '23]{Proceedings of the ACM Web Conference 2023}{May 1--5, 2023}{Austin, TX, USA}
\acmBooktitle{Proceedings of the ACM Web Conference 2023 (WWW '23), May 1--5, 2023, Austin, TX, USA}
\acmPrice{15.00}
\acmDOI{10.1145/3543507.3583875}
\acmISBN{978-1-4503-9416-1/23/04}

\begin{document}

\title{On Detecting Policy-Related Political Ads: \\ An Exploratory Analysis of Meta Ads in 2022 French Election}


\author{Vera Sosnovik}\authornote{Equal contribution.}
\affiliation{%
  \institution{Univ. Grenoble Alpes, CNRS,
Grenoble INP, LIG}
\country{France}
}

\author{Romaissa Kessi}\authornotemark[1]
\affiliation{%
  \institution{Univ. Grenoble Alpes, CNRS,
Grenoble INP, LIG}
\country{France}
}

\author{Maximin Coavoux}
\affiliation{%
  \institution{Univ. Grenoble Alpes, CNRS,
Grenoble INP, LIG}
\country{France}
}

\author{Oana Goga}
\affiliation{%
  \institution{LIX, CNRS, Inria, Ecole
Polytechnique, Institut Polytechnique de Paris}
\country{France}
}

%
%
%


\begin{abstract}

Online political advertising has become the cornerstone of political campaigns. The budget spent solely on political advertising in the U.S.\ has increased by more than 100\% from \$~700 million during the 2017-2018 U.S.\ election cycle to \$~1.6 billion during the 2020 U.S.\ presidential elections. Naturally, the capacity offered by online platforms to micro-target ads with political content has been worrying lawmakers, journalists, and online platforms, especially after the 2016 U.S.\ presidential election, where Cambridge Analytica has targeted voters with political ads congruent with their personality. 

To curb such risks, both online platforms and regulators (through the DSA act proposed by the European Commission) have agreed that researchers, journalists, and civil society need to be able to scrutinize the political ads running on large online platforms. Consequently, online platforms such as Meta and Google have implemented Ad Libraries that contain information about all political ads running on their platforms. 
This is the first step on a long path. Due to the volume of available data, it is impossible to go through these ads manually, and we now need automated methods and tools to assist in the scrutiny of political ads.

In this paper, we focus on political ads that are related to policy. Understanding which policies politicians or organizations promote and to whom is essential in determining dishonest representations. 
This paper proposes automated methods based on pre-trained models to classify ads in 14 main policy groups identified by the Comparative Agenda Project (CAP). We discuss several inherent challenges that arise. Finally, we analyze policy-related ads featured on Meta platforms during the 2022 French presidential elections period.  


\end{abstract}

%


\maketitle

\def\thefootnote{\arabic{footnote}}

\input{intro.tex}

\input{classification_taxonomy.tex}
\input{data.tex}

\input{model_configutation.tex}
\input{model_evaluation.tex}
\input{analysis_election.tex}

\input{lit_review.tex}

\input{conclusion.tex}

\section{Acknowledgements}
We thank the anonymous reviewers for their helpful comments.
This research was supported in part by the French National Research Agency (ANR) through the ANR-17-CE23-0014, ANR-21-CE23-0031-02, by the MIAI@Grenoble
Alpes ANR-19-P3IA-0003 and by the EU 101041223, 101021377 and 952215 grants.

\balance

\bibliographystyle{abbrv}
\bibliography{sample-base}

\input{Appendix.tex}


\end{document}

%% file: intro.tex
\section{Introduction}

Traditionally political parties have used manifestos to communicate the set of policies they announce they would implement if elected~
\cite{manifesto} and promoted their political agendas through mass media. With the emergence of online advertising platforms, online ads have become one of the main communication channels for political campaigners. 
During the 2020 US election cycle,  18\% of political marketing spending went to online advertising, compared to 3\%  during the 2016 election cycle \cite{forbes}. Moreover,  online advertising spending by parties increased from 24\% to 43\% of advertising budgets between the UK general elections of 2015 and 2017 \cite{electoralcomission}.



Besides the low cost, the key appeal of online micro-targeted advertising for political campaigners comes from the fact that they can communicate a more diverse set of information (than traditional mass media), and they can target subgroups of voters with information that is relevant to them.  
However, many researchers and civil societies are firing alarms that targeting technologies are also allowing the emergence of an ``industry of political influence''~\cite{Hankey-2019} where political advertisers can select very narrow groups of vulnerable people and tweak their messages to maximize their influence~\cite{ira_art}. 

The Cambridge Analytica scandal~\cite{Cmb-analyt} and the Russian's interference in the U.S.\ elections through online ads~\cite{IRA} has shaken to their core both online platforms as well as governments around the world. To curb such risks, both online platforms and regulators have agreed that researchers, journalists, and the civil society need to be able to scrutinize online political ads. Consequently, online platforms such as Meta and Google have implemented Ad Libraries that contain information about all political ads running on their platforms~\cite{google-ad-lib, metaadlibrary-5}. 
Better yet, ad libraries do not risk being discontinued. Thanks to the work of the European Commission in the Digital Services Act, access to ad libraries will be mandatory starting in 2024 for online platforms and search engines with more than 45 million monthly EU users~\cite{DSA}. This leads to new technical challenges: when faced with access to such large volumes of ad data (e.g., over 14M political ads in the U.S. and over 400k political ads in France on the Meta Ad Library)  it is impossible to go through these ads manually; hence, \emph{the new quest is to provide the public with methods and tools to assist in scrutinizing political ads.} 

In this paper, we focus on methods for \emph{detecting policy-related political ads}. There are a number of reasons why identifying policy-related political ads is important: (i) \emph{political communication}-- makes it possible to identify how political candidates and parties represent themselves and on which policies they focus their attention; (ii) \emph{mandate accountability}--check, once elected, whether elected officials respected the policy pledges they advertised during elections (accountability is central to democratic theory~\cite{thomson2011resolving}); 
(iii) \emph{influence on deliberation}--mandate theories assume that voters are rational and they decide for whom to vote based on a careful consideration of available information~\cite{Louwerse}. In practice, the deliberation process is more complex and is often based on emotions, convictions, and experiences~\cite{susser}. Policy-related ads are interesting in both ``rational voter'' and ``emotional voter'' models. Micro-targeting of policy-related ads could lead to some users being overly exposed to ads about specific policy issues (e.g., immigration), which might trigger strong emotions. In contrast, other voters might not get sufficiently exposed to any policy-related ads, which could lead to information incompleteness. 


For detecting policy-related ads, from a methodological perspective, we first need to decide what policy issues we should focus on and what is the right granularity (e.g., is ``economy'' too broad and should we consider ``climate change'' as an independent policy category?) While it is tempting to decide on a set of reasonable categories to detect, to robustly analyze policy-related ads across elections and countries, we need to rely on a solid, comprehensive, and stable theoretical basis. 
Luckily, two codebooks have been developed and polished by several groups of political scientists over several decades: the CAP  (Comparative Agenda Project) codebook~\cite{Baumgartner-2006} and the CMP (Comparative Manifesto Project) codebook~\cite{CMP}. The CAP codebook contains 28 main policy categories and 200 subcategories; while CMP contains 56 categories. 
The CAP codebook aims at capturing policy attention, and hence it aims at being comprehensive in the policy categories they propose~\cite{10.1093/oso/9780198835332.001.0001}. The CMP codebook aims at capturing political parties' ideological positions on a left-right scale, hence, focusing on ideological goals.   
In this paper, we use the CAP codebook as the underlying theoretical basis seems more suitable in the context of political micro-targeted ads.

For the analysis, we gathered more than 96k political ads from the Meta's Ad Library that appeared between 1 Jan and 14 June 2022  (sec.~\ref{data}). To gather labeled data, two experts annotated 431 ads with the relevant CAP categories. To complement this dataset, we used Prolific~\cite{prolific} and Qualtrics~\cite{qlt} to post assignments for annotating ads, and we gathered labels for 4 465 ads. We observe only fair agreement  (kappa>0.3) between Prolific users and experts. 
We show disagreement mainly happens on ads that are related to more than two policy categories (sec.~\ref{sec:label_quality}), hence, disagreement is linked to the text complexity of real-world ads. 

We implemented several machine learning (ML) models to classify ads in the relevant CAP categories based on both traditional supervised models and pre-trained language models based on BERT (sec.~\ref{classification}) that exploit as training data from CAP and annotations from Prolific users. Our best configuration is able to achieve a micro average F1 score of 0.60 over a balance test set (sec.~\ref{sec:model_eval}). The accuracy varies drastically depending on the policy category and ranges from a 0.19 F1 score for ``Social policy'' to a 0.78 F1 score for ``Environment''.  The differences are explained by the disagreement present in the training data and the labeling complexity of real-world ads.  



Finally, to show the practical usefulness of the classifier we developed, we analyze how policy attention varied across candidates and different demographic groups during the 2022 French Presidential election (sec.~\ref{frenchelections}). Overall, we see big variations in policy attention across demographic groups, with women over-targeted with ads about ``Health'', young users (ages 13-24) over-targeted with ads about ``Law and crime'' and users aged over 55 over-targeted with ads about ``Immigration''.  This kind of imbalance could reinforce gender and age stereotypes, and may deprive users from relevant information that might be important in their voting deliberation.

Through our study, we aim to provide a solid foundation for analyzing policy-related ads that combines knowledge from both political science and computer science research.  The challenges in solving the problem are diverse, and go from having the right codebooks for labeling, to having the right strategies to get high quality labels, and understanding which NLP algorithms are the most suitable for supporting such nuanced classification. All our datasets and code can be found at \url{https://www.lix.polytechnique.fr/~goga/datasets/policy_ads_www23.html}. During the French presidential elections, we have also developed a public web service to monitor the political ads send (\url{https://elections2022.imag.fr}).
The code of the web service can be found at \url{https://github.com/romaissalmh/elections2022}.

%% file: data.tex
\section{Data collections}
\label{data}

\subsection{Dataset of political ads}
We collected political ads featured on Meta's core advertising platforms during the 2022 French presidential election period (Jan 1st, 2022, to June 15, 2022). To do so, we built a data collection pipeline that, each day, retrieved the Meta's Ad Library daily report~\cite{reports_ad_lib}. This report contains information about advertisers (id and page name) who published ads, the number of ads, and the money spent.  We then used the advertisers' ids to retrieve all ads about social issues, elections, or politics using the Ad Library API. 


In total, we collected 91 865 unique political ads across 9 063 pages. We filtered only ads in French which lead to 76 886 ads. 
Since the Ad Library does not provide exact values of expenditures and impressions but intervals of values, we averaged these ranges and estimated the number of impressions to be around 4 billion (3 799 324 537) and 20 million euros spent (20 679 225). Appendix ~\ref{appendix} details the information provided by Meta for each ad. 

\subsection{Codebook for policy categories}
\label{taxonomy}

The literature in political sciences offers two noteworthy efforts for analyzing written political text: the Comparative Agendas Project (CAP) and the Comparative Manifesto Project (CMP). They are large-scale data collection efforts that gather and code information about the political processes of governments around the world based on the content of the texts. The effort is made by research groups in multiple countries from various disciplines and across multiple decades. 
These efforts have been allowing researchers, students, policymakers, and the media to study political trends over time and across countries. 

CMP's main goal is to archive and analyze the content of the electoral platforms of democratic countries from the end of the Second World War. CMP has proposed and is currently maintaining and updating a taxonomy that currently contains 54 categories~\cite{cmpcodebook}. The CMP codebook aims at capturing political parties' ideological positions on a left-right scale, hence, focusing on ideological goals. The CMP data collection classifies manifestos across multiple countries. For France, the CMP dataset contains 7 977 units of labeled text (long documents are split in text units that are labeled independently). 

CAP was created with the idea of tracking the attention of the government to particular policies. CAP creates a classification system that brings together a large number of political activities (e.g., bills, parliamentary debates, journalistic accounts) under a single theme with a taxonomy that counts 28 major topics (tab.~\ref{tab:nbAdPerCat}) and 250 subtopics  (e.g., waste is a sub-topic of the environment)~\cite{capcodebook}. The CAP codebook aims at capturing policy attention, and hence it aims at being comprehensive in the policy categories they propose. Contrary to CMP, CAP does not consider left-rights parties' positions and ideologies.  


In this paper, we decided to work with the CAP taxonomy. First, CAP's coding scheme focuses on the policy content and instruments, not political ideology, which we believe is more informative to study policy attention across demographic groups and candidates. Secondly, CAP's coding scheme aims to comprehensively cover topics of interest across countries (e.g., it does not miss important policy issues that might not exist in the U.S. but are essential in Vietnam). In contrast, the CMP codebook does not aim to be comprehensive. Finally, the CAP dataset has much richer data sources than CMP, which is only based on party manifestos. 

\vspace{2mm}
\noindent \textit{\capd:} For France, the CAP dataset contains 36 658 units of labeled text. The dataset contains text units from sources such as laws, government communications, decrees and bills, sentences from all major party manifestos for general legislative elections in France. Even if the labeled data is collected in a different domain than ours,  we use the \capd for training our classifiers. This type of training is called cross-domain transfer from a related domain and has been shown to work on other domains~\cite{5288526}. 



\subsection{Data labelling procedure}
To obtain labeled data, we hired human annotators to manually annotate political ads according to the~26 main CAP policy categories (tab.~\ref{tab:nbAdPerCat}). To account for political ads that are not policy-related we add an  ``Other'' category.
We encoded the survey using Qualtrics. Each survey contains one information page, followed by~1 page of \textit{task understanding tests}, then followed by~20 pages of texts of the ads to be labeled. Each ad page contained an ad's advertiser and text, followed by a list of~26 policy categories to choose from.
Going through a list of~26 policy categories is a hard task for workers. We pre-tested the survey with colleagues and workers to make the task more digestible. The survey version with the policy categories in bold, and short descriptions underneath was the most clear.  We gave Qualtrics a list of~5 000 texts of the ads, and we instructed it to randomly pick~20 texts to populate the survey at each instantiation. 

We then launched a study on Prolific where we redirected workers to the Qualtrics survey. The only requirement for workers was to be fluent in French. The survey took an average of~17 minutes to be completed. To determine the price to pay the workers, we took a reference payment of~7 pounds per hour (as suggested by Prolific). In total, we had~762 annotators.






We made sure that at least three different people annotated most ads to ensure the reliability of the assigned labels. Ads, that did not get three labels due to uneven Qualtrics' randomization mechanism, were deleted from the data set. We discarded all the answers from workers that took less than 4 minutes to complete the survey. As a result, the final set of labeled data consists of~4 465 ads. We selected the first three votes for ads with more than three annotations. Using these labels we created two labeled datasets:  

\vspace{1mm}
\noindent \textit{\vmd:} This dataset considers the voting majority. For each ad, we only keep the policy categories selected by two or three annotators. In case annotators agree on more than one policy category, we keep all of them. There are 3 784 political ads (out of 4 465) for which at least two annotators agreed on at least one policy category. We discard the ads for which there is no agreement from the dataset. 30\% of the ads are labeled with more than one policy category. Table~\ref{tab:nbAdPerCat} shows the number of ads per policy category in \vmd.  We selected 5 000 ads randomly. As a result the imbalanced distribution reflects the attention different policy categories were given during the French presidential period. We represent this dataset following the one-hot encoding, i.e. our data are in the form of a matrix M with:  $M_{ij}$ = 1 if >= 2 annotators chose theme j for ad i;  and $M_{ij}$ = 0 if not.



%
    
For the test dataset, to deal with the imbalance in the policy categories and ensure that we test on a reasonable proportion of each class, we randomly took from the \vmd~100 ad texts per category to form the test set. We ignored the categories for which we have less than 90 ad texts.  The test dataset contains 736 ads and we will call it the \test in the rest of the paper. 
We divided the rest of the data into training (2 160 ads) and validation (241 ads). 



\begin{table}[t]
\centering
\vspace{-3mm}
\caption{Number of labeled ads on Prolific per policy category.}
\vspace{-3mm}

\footnotesize{
\begin{tabular}{lr}
    \toprule
Policy category & Number of ads   \\
\midrule
\rowcolor{color1} \textcolor{white} {Environment} & \textcolor{white}{683}    \\
\rowcolor{color1} \textcolor{white} {Human rights} & \textcolor{white} {623} \\
\rowcolor{color1} \textcolor{white} {Cultural policy} & \textcolor{white} {469} \\
Others & 403 \\
\rowcolor{color1} \textcolor{white} {Health} & \textcolor{white} {374}\\
\rowcolor{color1} \textcolor{white} {Social policy} & \textcolor{white} {340}\\
\rowcolor{color1} \textcolor{white} {Energy} & \textcolor{white} {318}\\
\rowcolor{color1} \textcolor{white} {Government operations} & \textcolor{white} {311}\\
\rowcolor{color1} \textcolor{white} {International affairs} &  \textcolor{white} {258}\\
\rowcolor{color2} Work and employment & 189\\
\rowcolor{color1} \textcolor{white} {Macroeconomic policy} & \textcolor{white} {185}\\
\rowcolor{color2} \textcolor{black} {Education} & \textcolor{black} {146}\\
\rowcolor{color2} \textcolor{black} {Justice and criminality} & \textcolor{black} {136}\\
\rowcolor{color1} \textcolor{white} {Economic regulations} & \textcolor{white} {132}\\
\rowcolor{color2} {Urban and territorial policies} &  115\\
\rowcolor{color2}  {Immigration} &  96\\
Transport & 69\\
Agriculture & 69\\
Technology and communication & 64\\
Defense & 54\\
Religion & 52\\
Foreign trade & 40\\
Sports & 38\\
Risk and natural disasters & 22\\
Fires and accidents & 3\\
Public domain and water management&0\\
Local and regional policy&0\\
Obituary&0\\
\bottomrule
\end{tabular}}
\label{tab:nbAdPerCat}
\vspace{-6mm}
\end{table}

\vspace{2mm}
\noindent \textit{\disd:} To take into account all annotations, we create a second dataset that contains the distribution of annotations on policy categories. 
Prior research~\cite{Fornaciari-2021} has shown the empirical benefit of predicting soft labels, i.e.\ probability distributions on annotators' labels, as an auxiliary task to take into account annotators' disagreement.
The \disd contains all the 4 465 previously annotated advertisements but considers soft labels. The matrix representation is done as follows:$M_{ij}$ =  0.3 if when 1 annotator selected category $j$ for ad $i$; 
      $M_{ij}$ = 0.6 when 2 annotators selected  category $j$ for ad $i$; 
      $M_{ij}$ = 1 if when 3 annotators selected category $j$ for ad $i$.
We use the \disd for training and validation, but not for testing. We split \disd in train set (4 000 ads) and validation set (370 ads).  

\subsection{Analyzing annotation quality}
\label{sec:label_quality}
While we took several steps to make the labeling task as easy as we could for workers, we still observe a lot of disagreement on the policy categories chosen by different workers: on 16\% of the ads annotators did not agree on \emph{any} policy category. One reason for the observed disagreement could be due to the limited comprehension of the assignments by workers that try to perform tasks as fast as possible. 
Another reason might be the intrinsic difficulty of the task, i.e., even experienced annotators with a lot of time on their hands would disagree on the policy category \cite{gone_fishing}.
To assess the quality of the Prolific annotations, two expert annotators (the Ph.D. students working on the project), annotated independently ~50\% of the \test (431 ads). The two experts disagreed on 10\% of the ads. After discussions and reading the codebooks several times, the expert annotators agreed on at least one policy category for the ads they initially disagreed on. 
In what follows, we refer to their annotations as \emph{gold labels} and the corresponding dataset as \testgs. In \testgs we only keep the policy categories the two expert annotators agreed on.

Inter-annotator agreement measures are widely used to quantify the reliability of data annotations~\cite{artstein-poesio-2008-survey}, or to establish an upper-bound on a systems' performance~\cite{amidei-etal-2018-rethinking}.
Table~\ref{tab:agreemntReportGold} shows the pair-wise Cohen Kappa between the final labels of Prolific workers (the \vmd) and the final labels of experts. There is a fair agreement (>0.3) for all categories, but a substantial agreement (>0.6) only for five categories. 
We observed by looking at the ads on which there is disagreement that they tend to have more labels from either experts or Prolific workers.  To validate this intuition, Table~\ref{tab:agreemntReportGold} shows the inner-annotator agreement separately for ads with 1-2 categories (208) and ads with more than 2 categories (223).  We see that for ads labeled with only 1 or 2 policy categories the agreement is substantially higher than for ads labeled with more than 2 categories.  
Hence, ads that relate to multiple policy categories are more confusing and lead to disagreement. However, we do observe substantial and almost perfect agreement on the rest of the ads.


\begin{table}[t]
\centering
\caption{Agreement between gold labels and Prolific labels for all ads, for ads with 2 or less policy categories and for ads with more than 2 policy categories.}
\vspace{-3mm}
\footnotesize{
\begin{tabular}{lccc}
    \toprule
Ads  &all & 1-2 policy cat. &>2 policy cat.\\
  \midrule
International affairs & 0.62 & 0.68&0.47\\
Energy & 0.75&0.88&0.33 \\
Government operations & 0.58&0.67&0.31\\
Cultural policy & 0.68&0.81&0.22\\
Social policy & 0.38&0.44&0.19\\
Health&0.68&0.8&0.41\\
Human rights & 0.49&0.72&0.12\\
Environment &0.61&0.73&0.27\\
Economy&0.34&0.47&0.04\\
\bottomrule
\end{tabular}}
\label{tab:agreemntReportGold}
\vspace{-5mm}
\end{table}

To dig deeper into disagreement, Table~\ref{tab:test-classificationReportGold} (appx.~\ref{appendix}) shows the classification ratio assuming that our golden labels are the real labels and the Prolific labels are predictions. 
The ``Social policy'' category has the highest number of false positives (small precision), while the  ``Economy'' category has both high false positives (small precision) and high false negatives (small recall). 
Table~\ref{tab:exampl_disagr} (app.~\ref{appendix}) shows three examples of ads for each policy category that are false positives and false negatives. On reason for false positives is because Prolific workers interpret more loosely the 26 policy categories. For example, the ad: ``\emph{The situation on the Ukraine - Russia border is more than tense. Far be it from me to think that my opinion on this subject is particularly relevant. However, I am convinced that by turning to past history, we can try to shed light on certain points of this burning issue.}'' was labeled as being related to ``Economy'', ``Human rights'' and ``International affairs'' by Prolific workers but was only labeled as ``International affairs'' by experts. False negatives seem to happen when experts label ads with multiple categories, while Prolific workers label the ads with only a subset of categories. This might happen because Prolific workers try to limit the time they spend to label an ad and once they find a few relevant categories they go to the next ad. 

To check incomprehension in the task, we look at differences in the confusion matrices of Prolific workers and experts. The confusion matrix of Prolific workers' labels (fig.~\ref{fig:heat_map_before_gold}, appx.~\ref{appendix}) shows that a higher number of ads is labeled as both ``Energy'' and ``Environment'' as well as ``Social policy'' and ``Human rights'', while the confusion matrix of gold labels (fig.~\ref{fig:heat_map_before_gold}, appx.~\ref{appendix}) displays a lower intersection. 
Hence, one reason for disagreement is that the some workers do not see clearly enough the difference between  ``Energy'' and ``Environment'' as well as  ``Social policy'' and ``Human rights''.






%% file: model_configutation.tex
\section{Classification models}
\label{classification}

 
The task of detecting ads related to different policy categories corresponds to a \emph{multi-label classification problem}. In multi-label classification, the training set consists of instances associated with a set of labels. In our case, we assume that an ad may refer to multiple policy categories. We built several classifiers where we tested different training datasets and hyperparameter configurations of both traditional supervised methods and recent methods based on large pre-trained language models \cite{devlin-etal-2019-bert}. For each classifier configuration, we build one \emph{multi-label classifier}. We also tried One vs.\ All methods (i.e., having one classifier for each policy category instead of one multi-class classifier); however, it led to significantly worse results, probably because One vs.\ All methods do not consider any underlying correlation between the classes. 



\vspace{2mm}
\textbf{Training data.}
Ideally, we would have large amounts of labeled data annotated by domain experts. Due to the unavailability of such dataset, we exploit training data that is less clean but easier to collect. Building policy detection algorithms without spending months to collect gold labels is practical in the real-world, especially if we want detection algorithms that work across languages and elections. 
We instantiated three training sets based on the dataset described in Section~\ref{data}: \vmd, \capd, and \disd.

\vspace{1mm}
\noindent  \textit{Data prepossessing:}
Prior to training, we remove links and emojis from the text of the ads. In addition, for supervised models, we also delete stop-words and punctuation signs.


\vspace{1mm}
\noindent \textit{Data augmentation:} We use
a classical data augmentation approach based on back-translation to increase the training set.
Back translation consists in automatically translating an input text into another language, and then translating it back to French. The resulting text should be a paraphrase of the original text (if it is not identical to it) and can be used as a synthetic training example with the same label as the original text.
We apply back translation with 40\% of the ads from the train set for each category as a pivot language, and augment the training datasets to 2 542 examples (from 2 160 examples).

\vspace{2mm}
\textbf{Supervised models.}
 \label{sec:bow}
As a baseline, we chose two popular supervised models for our task: SVM \cite{cortes1995support} and Random Forests \cite{breiman2001random}. To convert the words of our data to numeric representations, we tried three vectorization techniques such as bags of words, hashing vectoriser, and TF/IDF. The last one, being the one that outperformed the others, was chosen for the rest.
We used grid search to calibrate the hyperparameters of SVM and Random Forest using 10-fold cross validation.

\vspace{2mm}
\textbf{Pre-trained language models.}
\label{sec:bert}
We use classifiers based on pre-trained language models since they are the current state of the art in text classification~\cite{devlin-etal-2019-bert,DBLP:journals/corr/abs-1907-11692}. 
Large language models such as BERT~\cite{devlin-etal-2019-bert} are pre-trained on a very large amount of raw unlabeled texts (typically tens of Gb) with a self-supervised objective. They provide good-quality representations for words and sentences.
Pre-trained language models have the advantage of working well with limited labeled examples thanks to the richness of the sentence representations they provide.
For French, the language we are dealing with, there exist two main pre-trained models: \textbf{CamemBERT}~\cite{martin-etal-2020-camembert} and \textbf{FlauBERT}~\cite{le-etal-2020-flaubert-unsupervised}. Both models are based on the BERT architecture ~\cite{devlin-etal-2019-bert}, and have been trained with a masked language modeling (MLM) objective. 
CamemBERT is trained on 138 Gb of textual data crawled from the web, and FlauBERT is trained with 71 Gb of data from diverse sources, including crawled data and Wikipedia.





In order to perform classification, we use the pre-trained language model to extract a vector representation for the input text, and feed this representation to a linear classification layer with a sigmoid activation. We obtain a vector $\mathbf p \in [0,1]^9$, such that 
$\mathbf p_l = p(l|\text{ad})$ interprets as  the probability of label $l$ given an ad: the probability of each label is modeled independently.
In other words, each policy category has its own binary classifier which is siutable for our multi-label classification.
Then, for a given ad, we assign it all labels $l$ for which $p(l|\text{ad}) > t_l$ where $t_l$
is a threshold. 
As a result, a single text can be assigned any number of labels (0, 1, 2 or more).
A typical threshold is to use 0.5 for every category.
The threshold value can also be used to control the trade off between higher recall and higher precision (sec.~\ref{sec:final_model}).


\vspace{2mm}
\noindent \textit{Training.}
We optimize a binary cross-entropy loss function to train the model, for 4~epochs and a learning rate of $2\text{e-}5$.
For better convergence, we used a linear-decreasing learning rate during optimisation and a batch size of~8.
Our implementation uses the \texttt{transformers} library \cite{huggingface} for FlauBERT and CamemBERT pre-trained models.

%% file: model_evaluation.tex
\section{Model evaluation}
\label{sec:model_eval}

In this section, we evaluate the models described in sec.~\ref{classification}
in order to select the best model for the next part of our study (sec.~\ref{frenchelections})
and we provide an analysis of their behaviour to better understand the limitations.
In particular, we assess the effects of the classification algorithm, and the training dataset.

\vspace{1mm}
\textit{Evaluation sets.}
Due to the category imbalance in the data (sec.~\ref{data}), some policy 
categories are very infrequent.
Therefore, in order to make evaluation more reliable, 
we build and test classifiers on a subset of the data with the~9 policy categories
with more than~200 labeled ads, namely: \textit{environment}, \textit{international affairs}, \textit{energy}, \textit{civil rights}, \textit{government operations}, \textit{health}, \textit{social policy}, \textit{cultural policy}, and \textit{economy} (which includes foreign trade, macroeconomic policy and economic regulations).
We call this evaluation dataset \testn. It contains 736 ads
whose labels are obtained by a majority vote from Prolific workers.
In addition, we evaluate models on \testgs, the subset of \testn for which we have labels provided
by domain experts (sec.~\ref{data}). \testgs contains 431 ads.
%
%
After performing model selection on \testn and \testgs, we retrain the best classifier over a training set with~14 policy categories by adding categories that have more (or close to) 100 labeled ads. We will base our study in sec.~\ref{frenchelections} on this retrained model.

\vspace{1mm}
\textit{Evaluation metrics.} For each of our experiments, we report traditional 
evaluation metrics for text classification, namely: precision, recall and F$_1$ score for
each category, as well as a micro-average across the whole test set of these metrics.



\begin{table}[t]
\centering
\caption{Accuracy across models over \testn. The tables presents the micro-averages of precision, recall and F1 scores.}
\vspace{-3mm}
\footnotesize{
\begin{tabular}{lrrr}
    \toprule
    & Precision & Recall & F-1\\
    \midrule
    SVM & 0.45 & 0.40 & 0.52 \\
    Random Forest & 0.39 &  0.33 & 0.46 \\
    FlauBERT & \textbf{0.79} & \textbf{0.59} & \textbf{0.68} \\
    CamemBERT & 0.72 & 0.61 & 0.66 \\
    \bottomrule
    \end{tabular}
    }
\label{tab:classificationResultGlobal}
\vspace{-5mm}
\end{table}

\subsection{Results}

\paragraph{Comparing classification algorithms}
We first train the four models we used (i.e., SVM, Random Forest, FlauBERT and CamemBERT) on \vmd and we report the accuracy of the best configuration of each classifier, as selected by 10-fold cross validation.
We present the results of their predictions on \testn in Table~\ref{tab:classificationResultGlobal}.
As expected, FlauBERT and CamemBERT outperform SVMs and Random Forests by a large margin, and obtain F$_1$ scores over 0.65.
This is in line with current research in NLP: the pre-training on massive amounts of unlabeled data makes language models able to adapt quickly to a downstream task, even when the size of the training set is small. In what follows, we settle on the FlauBERT-based classifier, that slightly outperformed the CamemBERT-based classifier.

\vspace{2mm}
\textit{Comparing training sets.}
Models whose results are reported in Table~\ref{tab:classificationResultGlobal} were trained on \vmd.
However, recall that we also have \capd, a dataset which contains a different type of documents
(sec.~\ref{data}) but is nevertheless much larger (25.4k labeled examples).
We hypothesize that the size of this dataset may compensate for the discrepancy in terms of types of documents, and that the resulting model would generalize well on our test set, achieving cross-domain knowledge transfer.

We present the results of the models trained on \capd in Table~\ref{tab:classificationResultsoftlabels}.
Unfortunately, the hypothesis turned out wrong: when trained on \capd, the FlauBERT-based classifier
only achieved 0.13 F$_1$.
This could be due to the domain discrepancy between the political ads from Meta and the documents in \capd, in terms of vocabulary distribution or length (the average length of a CAP document is 36 tokens, whereas it is an 63 tokens for an ad).
Moreover, in \capd, each document has a single label, whereas in the evaluation set, an ad may have several labels, leading to a distribution shift between the train and test set which might confuse the model.





\begin{table}[t]
\centering
\caption{Accuracy across training datasets. Comparison of FlauBERT's accuracy trained with \capd, \vmd (majority vote labels), and \disd (soft labels).}
\vspace{-3mm}
\footnotesize{
\begin{tabular}{llrrr}
    \toprule
    & Training set & Precision & Recall & F-1\\
    \midrule
    FlauBERT & \capd & 0.14 & 0.11 & 0.13 \\
    FlauBERT & \vmd & 0.79 & 0.59 & 0.68\\
    FlauBERT & \disd  & 0.79& 0.60& 0.68\\
    \bottomrule
    \end{tabular}
    }
\label{tab:classificationResultsoftlabels}
\vspace{-5mm}
\end{table}

Prior work has shown that soft labels might help the classifier \cite{plank-etal-2014-learning,martinez-alonso-etal-2016-supersense,Fornaciari-2021}.
Indeed, disagreement between annotators is not only due to noise, but can also
contain an important signal.
For example, if two categories are systematically prone to disagreement, they might overlap
partially in their definition. 
This signal can be exploited by a classifier by weighting the labels in the training data by the proportion of annotators who chose a specific label for a give example, as an indication of uncertainty for the model.

We investigate whether modelling annotator disagreement helps in our case by training the FlauBERT-based classifier on \disd.
The resultants are presented in Table~\ref{tab:classificationResultsoftlabels}:
the training with soft labels does not improve upon training on majority-voted labels.

\vspace{2mm}
\textit{Results per category.}
Table~\ref{tab:test-9cat-th} illustrates the precision, recall and F1 score of FlauBERT across the~9 policy categories. The support is the number of texts in a specific category in the test set. 
We observe that some policy categories such as \textit{environment, energy} and \textit{cultural policy} are well detected, whereas the accuracy is much lower for ads related to \textit{social policy}.
Overall, these categories with high accuracy correspond to those with a higher agreement between annotators (tab.~\ref{tab:agreemntReportGold}), and conversely \textit{social policy} 
and \textit{economy} have the lowest agreement.
Indeed, a low agreement indicates both that the annotations are less reliable, and that the category is harder to detect.


\begin{table}[t]
\caption{Accuracy across policy categories using FlauBERTtuned over \testn.}
\footnotesize{
\begin{tabular}{lrrrr}
    \toprule
    &Prec. & Rec. & F-1 & Support  \\
    \midrule
    International affairs & 0.81 & 0.60 & 0.69&100 \\
    \rowcolor{bittersweet}Energy & 0.93 &0.68&0.79&100 \\
    Government operations & 0.65 & 0.43&0.52&105\\
    \rowcolor{bittersweet}Cultural policy & 0.84 & 0.83&0.83&109\\
    \rowcolor{ballblue}Social policy &0.76& 0.19&0.30&102\\
    Health&0.86&0.73&0.79&102\\
    Human rights& 0.67&0.47&0.55&125\\
    \rowcolor{bittersweet}Environment&0.81&0.80&0.81&150\\
    Economy&0.75&0.49&0.59&102\\
    \midrule
    micro avg&0.79&0.59&0.68&995\\
    samples avg&0.72&0.64&0.66&995\\
    \bottomrule
\end{tabular}}
\label{tab:test-9cat-th}
\vspace{-4mm}
\end{table}

\subsection{Evaluation of the final model}
\label{sec:final_model}
The previous section showed that the overlap in content between policy categories has a negative impact on the achievable accuracy. 
In this section, we look at how accuracy changes when we consider more policy categories. 
Here, we train and test classifiers using~14 policy categories for which we have more than 100 ads. 
We add the following policy categories: \textit{education, justice and crime, work and employment, urban and territorial policies}, and \textit{immigration.} 
Unfortunately, we do not have enough labeled data to add the 12 other categories.
Table~\ref{tab:test-14cat-th} shows the results across the~14 policy categories. 
The table shows that even for policy categories such as \textit{immigration} and \textit{urban and territorial policies} for which we have less than 200 ads, the classifier is able to achieve F1 scores over 0.5. 
The table also shows that the accuracy of the initial~9 policy categories slightly drops. 
Indeed, the additional categories make the task harder.
A higher number of categories also leads to higher potential overlap between categories.


\begin{table}[t]
\centering
\caption{Accuracy across policy categories using FlauBERTtuned over the \testf.}
\vspace{-2mm}
\footnotesize{
\begin{tabular}{lrrrr}
    \toprule
  &Prec. & Rec. & F-1 & Support  \\
  \midrule
International affairs & 0.68 & 0.49 & 0.57&102 \\
Energy & 0.92 &0.59&0.72&100 \\
Immigration&0.71&0.50&0.59&30\\
Law and crime&0.83&0.29&0.43&35\\
Government operations & 0.64 & 0.41&0.50&105\\
Cultural policy & 0.83& 0.71&0.76&110\\
Social policy & 0.85& 0.11&0.19&104\\
Urban and territorial policies&0.61&0.45&0.52&31\\
Health&0.85&0.67&0.75&101\\
Labor and employment & 0.59&0.61&0.60&36\\
Human rights& 0.63&0.51&0.57&134\\
Education&0.88&0.23&0.36&31\\
Environment & 0.76& 0.79&0.78&150\\
Economy&0.89&0.23&0.37&103\\
\midrule
micro avg&0.75&0.50&0.60&1172\\
samples avg&0.64&0.56&0.58&1172\\
\bottomrule
\end{tabular}
\label{tab:test-14cat-th}}
\vspace{-4mm}
\end{table}


\vspace{1mm}
\textit{Controlling the precision-recall trade off.}
For our case study, the precision is more important than the recall--it is more important not to mislabel ads with the wrong policy category than to miss some ads that are related to a policy category. 
Given this preference, instead of using the same threshold for each category (i.e., 0.5), we select a different threshold for each policy category. 
To get the appropriate threshold for each category we performed \textit{threshold optimization} as a fine-tuning step. The definition of the threshold is done during the validation phase by maximizing precision over recall. We look for thresholds that give a precision close to 85\% with the highest possible recall. 

%
%

Table~\ref{tab:test-14cat-tho} (appx.~\ref{appendix}) presents the precision and recall of FlauBERT with~14 policy categories using different thresholds.  Note that the precision is not always close to 0.85 since the thresholds have been defined on validation data and not on test data.  In the next section we use this model for label prediction. 

%% file: analysis_election.tex
\section{Case Study: Policy attention in the 2022 French election Ads}
\label{frenchelections}

Political scientists and analysts have long been interested in policy attention dynamics across countries and elections~\cite{elumpr}. However, most studies have analyzed policy attention through manual annotations of various sources of texts such as political parties manifestos, mass media, and senate hearings. 
 As a case study, to show the practical usefulness of the classifier we developed in the previous section, we analyze how policy attention varied across candidates and different demographic groups during the 2022 French Presidential election (held in two rounds: 10 April and 24 April). 
We applied the FlauBERT model for the analysis with different thresholds on the 76 067 political ads we collected from Meta's Ad Library that ran between 1 January 2022 and 15 June 2022. FlauBERT model with different thresholds ensures high precision, hence, being confident that all the ads labeled about a specific policy are correct. However, the recall varies across policy categories from 0.16 to 0.75. Hence, we cannot detect \emph{all} ads corresponding to every policy category. For this section, this is not problematic as our analysis compares policy attention in different demographic groups and across presidential candidates, and a low recall should count equally in all groups.\footnote{Different recalls for different policy categories will be problematic if the goal of the analysis would be to determine the policy categories that most attention.} 
Out of the 76 067 political ads, our model predicted at least one policy category for 59 718 ads. Moreover, for 6 531 ads the model predicted more than one policy category and ads had in median 1 policy category. 


\subsection{Policy attention and presidential candidates}

We analyze both policy attention in ads coming from the official accounts of presidential candidates and their corresponding political parties and ads that do not necessarily come from official accounts but mention a candidate's name. Remember that on Meta, anyone can be an advertiser and send political ads if they prove they reside in the same country where the ads are targeted.

There were 12 candidates in the election, and we manually found all official corresponding accounts. 
In France, the law prohibits, in the six months preceding an election, the use for electoral propaganda purposes of any commercial advertising process by the press or by any means of audiovisual communication~\cite{lawElectoral}. Despite the law, we observed 321 ads (corresponding to 23 443 021 million ad impressions) coming from several official accounts of presidential candidates posted from 1 Jan to 24 Apr. We saw Emmanuel Macron's party ``En Marche'' circumventing this prohibition by financing a few days before the elections ``register to vote'' ads on Facebook targeting users ages 18-24 posted on the page ``La France aux urnes''.\footnote{https://www.facebook.com/la.france.aux.urnes.2022} In addition, Eric Zemmour and Marine Le Pen (two prominent right-wing extremists) also sponsored political ads encouraging users to join their party or support them through donations. To see the content of these ads, please check our Election analysis server at \url{https://elections2022.imag.fr}.   

Secondly, we identified 1598 ads that mention one of the top three presidential candidates according to votes in round 1: 1 050 mention Emmanuel Macron, 406 mention Marine Le Pen, and 142 mention Jean-Luc M\'elenchon. Table~\ref{tab:candidates_policy} shows the policy attention of ads corresponding to different presidential candidates. To measure policy attention, we collected information from Meta's Ad Library about the number of ad impressions (i.e., the number of users that saw the ad) of each ad in our dataset (sec.~\ref{data}). Hence, we summed up the ad impression information for all ads mentioning a particular candidate and labeled with a particular policy category. 
The table shows that many ads mentioning the candidates address ``Government operations'' which makes sense since this category describes everything related to the elections and the country's state. The distribution of ad impressions across the other policy categories is uneven across candidates. The majority of the ads that mention Macron discuss ``International affairs''. This can be justified by the strong involvement of the French president in the war between Ukraine and Russia. In contrast, most ads mentioning Le Pen discuss ``Health'' and most ads mentioning M\'elenchon (besides ``Government operations'') discuss ``Economy''. Understanding how candidates represent themselves and on which policies they focus their attention, and how the large public talks about the candidates is important for mandate accountability and understanding how democracies evolve.

\begin{table}[]
\caption{Distribution of ads impressions by policy category in ads mentioning different presidential candidates.}
\vspace{-3mm}
\footnotesize{
\begin{tabular}{lccc}
\toprule
                             & E. Macron & M. Le Pen & J-L.  M{\'e}lenchon \\
                           \midrule
International affairs        &  \cellcolor{bittersweet}{29.14\%}        & 7.37\%        & 1.77\%             \\
Energy                       & 0.48\%          & 0.00\%        & 0.00\%             \\
Immigration                  & 0.26\%          & 0.11\%        & 0.00\%             \\
Law and crime      & 0.34\%          & 0.00\%        & 0.00\%             \\
Gouvernement operations      &  \cellcolor{bittersweet}{30.69\%}         &  \cellcolor{bittersweet}{16.00\%}       &  \cellcolor{bittersweet}{63.02\%}              \\
Cultural policy              & 3.27\%          & 0.06\%        & 0.00\%             \\
Social policy                & 0.72\%          & 0.07\%        & 1.47\%             \\
Urban and territorial policy & 0.30\%          & 0.09\%        & 0.04\%             \\
Health                       & 2.75\%          &  \cellcolor{bittersweet}{49.03\%}       & 1.51\%             \\
Work and employment          & 0.73\%          & 0.07\%        & 0.00\%             \\
Human rights                 & 4.39\%          & 0.31\%        & 1.26\%             \\
Education                    & 0.95\%          & 0.37\%        & 0.00\%             \\
Environment                  & 10.52\%         &  \cellcolor{bittersweet}{14.84\%}      & 0.00\%             \\
Economy                      &  \cellcolor{bittersweet}{15.46\%}         & 11.68\%       &  \cellcolor{bittersweet}{30.91\%} \\
\bottomrule
\end{tabular}}
\label{tab:candidates_policy}
\vspace{-5mm}
\end{table}

\subsection{Policy attention across demographic groups}

Meta's Ad Library provides information on the demographic distribution of people reached by every political ad. In Table~\ref{tab:demog_policy}, we use this information to study policy attention across demographic groups by investigating what are the demographic groups that are over/under targeted by the different policy categories. Each cell represents the proportion of ad impressions of ads related to a particular policy categories that have been viewed by a particular demographic group.  
The first line of the table (i.e. Population) represents the demographic distribution of all ad impressions in French that have at least one predicted policy. We use this as a baseline to identify over-exposure (in red) and under-exposure (in blue). 
A few interesting observations we see in the table:
\begin{trivlist}
\item (1) Women are under-exposed (compare to men) to ads talking about ``Energy'' and ``Economy'' and they are over-exposed to ads about ``Immigration'', ``Social policy'', and ``Health''.

\item (2) Users aged 18-24 are under-targeted to ads about ``International affairs'', while users over 65 are over-targeted. 
\item (3) Users aged 18-34 are under-targeted to ads about ``Immigration'', while users over 45 are over-targeted. 
\item (4) Users aged 13-24 are severely over-targeted with ads about ``Law and crime''. 
\item (5) ``Cultural policy'', ``Social policy'', ``Economy'', and ``'Human rights'' are pretty evenly distributed across demographic groups.
\item (6) Users aged over 55 are over-targeted with ads about ``Health''.
\item (7) Users aged 18-24 are over-targeted with ads about ``Work and employment''. 
\end{trivlist}
Overall, we do see large variations in policy attention across demographic groups.  This kind of imbalance may not be beneficial as it could reinforce gender and age stereotypes, and may deprive certain users from relevant information that might be important in their voting deliberation. Who received an ad depends on both the advertiser that can specify to which gender and age groups they want to send their ad, but also the ad optimization algorithms employed by Meta~\cite{10.1145/3359301} that optimize ad deliver. To better understand who is responsible for the imbalance in policy attention, it is necessary that ad platforms provide more transparency about the demographics chosen by the advertiser, and the demographics of the user the ad actually reached. 

Table~\ref{tab:region_policy} (appx.~\ref{appendix}) shows the distribution of ad impressions across regions and policy categories.  
There are differences between regions, we see that around 60\% of all ads about ``Justice and criminality'' were shown to the people from the Ile-De-France region; and more than 27\% of the total ads about ``Urban policies and territories'' were shown in Auvergne-Rhône-Alpes.  


\begin{table*}[t]
\caption{Distribution of ad impression across demographic groups and policy categories. $^*$ represents the demographic distribution of impressions for all ads in French that have at least one predicted policy. Over-exposure (in red) and under-exposure (in blue).}
\vspace{-3mm}
\footnotesize{
\begin{tabular}{l|l|l|l|l|l|l|l|l|l}
\toprule
                             & \multicolumn{2}{c|}{\textbf{Gender}}  & \multicolumn{7}{c}{\textbf{Age}} \\
                             \hline
                             & Female  & Male    & 13-17   & 18-24   & 25-34   & 35-44   & 45-54   & 55-64   & 65      \\
\midrule
Population$^*$ (baseline)                  & 53.94\% & 46.06\% & 2.68\%  & 14.24\% & 22.79\% & 18.33\% & 15.14\% & 13.12\% & 13.52\% \\
\midrule
International affairs        & 53.64\% & 46.36\% & 0.18\%  &     \cellcolor{ballblue}{6.26}\%  &     {22.22\%} & 19.83\% & 16.64\% & 15.89\% & \cellcolor{bittersweet}{18.98\%} \\
Energy                       & \cellcolor{ballblue}{35.61\%} & \cellcolor{bittersweet}{64.39\%} & 0.01\%  &  \cellcolor{ballblue}{1.74\%}  & 25.76\% & \cellcolor{bittersweet}{34.49\%} & \cellcolor{bittersweet}{27.35\%} & \cellcolor{ballblue}{7.88\%}  &  \cellcolor{ballblue}{2.77\%}  \\
Immigration                  &  \cellcolor{bittersweet}{65.65\%} &  \cellcolor{ballblue}{33.35\%} & 0.18\%&  \cellcolor{ballblue}{4.6\%} & \cellcolor{ballblue}{16.68\%}   & 17.06\% & \cellcolor{bittersweet}{18.88\%} & \cellcolor{bittersweet}{20.57\%} & \cellcolor{bittersweet}{22.02\%} \\
Law and crime      & 51.88\% & 48.12\% &  \cellcolor{bittersweet}{28.57\%} & \cellcolor{bittersweet}{24.52\%} & \cellcolor{ballblue}{8.46\%}  & \cellcolor{ballblue}{10.97\%} & \cellcolor{ballblue}{9.3\%}   & \cellcolor{ballblue}{8.33\%}  & \cellcolor{ballblue}{9.85\%}  \\
Government operations      & 53.35\% & 46.65\% & 0.92\%  & \cellcolor{bittersweet}{30.26\%} & \cellcolor{bittersweet}{27.32}   &  \cellcolor{ballblue}{13.83\%} & \cellcolor{ballblue}{9.7\%}   & \cellcolor{ballblue}{8.7\%}   & \cellcolor{ballblue}{9.23\%}  \\
Cultural policy              & 51.53\% & 48.47\% & 3.17\%  & 16.18\% & 23.32\% & 17.97\% & 15.63\% & 12.21\% & 11.52\% \\
Social policy                &  \cellcolor{bittersweet}{65.34\%} & \cellcolor{ballblue}{34.66\%} & 0.93\%  & 13.65\% & 19.74\% & 16.27\% & 14.51\% & 16.15\% & \cellcolor{bittersweet}{18.74\%} \\
Education                    & 59.19\% & 40.81\% & \cellcolor{bittersweet}{11.4\%}  & \cellcolor{bittersweet}{24.2\%}  & \cellcolor{ballblue}{16.6\%}  & \cellcolor{ballblue}{13.8\%}  & \cellcolor{ballblue}{9.67\%}  & 10.26\% & 14.79\% \\
Environment                  & 50.89\% & 49.11\% & \cellcolor{ballblue}{1.95\%}  & \cellcolor{ballblue}{10.78\%} & 25.52\% & 21.91\% & 17.04\% & 12.17\% & \cellcolor{ballblue}{10.64\%} \\
Health                       &  \cellcolor{bittersweet}{68.45\%} & \cellcolor{ballblue}{31.55\%} & \cellcolor{bittersweet}{4.16\%}  & \cellcolor{ballblue}{8.62\%}  & \cellcolor{ballblue}{14.18\%} & 18.12\% & 17.12\% &  \cellcolor{bittersweet}{18.42\%} &  \cellcolor{bittersweet}{19.37\%}\\
Economy                      & \cellcolor{ballblue}{44.24\%} &  \cellcolor{bittersweet}{55.76\%} & \cellcolor{ballblue}{0.01\%}    & 12.85\% & 22.25\% & 19.36\% & 15.67\% & 15.53\% & 14.34\% \\
Human rights                 & 59.31\% & 40.69\% & \cellcolor{bittersweet}{8.74\%}  & 16.18\% & 20.09\% & 16.52\% & 13.54\% & 12 \%   & 12.13\% \\
Work and employment          &  \cellcolor{bittersweet}{61.27\%} & \cellcolor{ballblue}{38.73\%} & 0.84\%  &\cellcolor{bittersweet}{27.49\%} & 19.03\% & \cellcolor{ballblue}{12.60\%} & 14.79\% & 12.88\% & 12.37\% \\
Urban and territorial policy & 51.09\% & 48.91\% & 3.63\%  & \cellcolor{ballblue}{6.32\%}  & \cellcolor{ballblue}{14.62\%} & \cellcolor{ballblue}{15.90\%} & 15.60\% &  \cellcolor{bittersweet}{17.65\%} &  \cellcolor{bittersweet}{26.29\%} \\
\bottomrule
\end{tabular}
}
\label{tab:demog_policy}
\vspace{-3mm}
\end{table*}


%% file: lit_review.tex
\section{Related Work}

\paragraph{Works on online political ads} 
A few early works have used \emph{manual labeling} to encode political ads according to various characteristics and analyze the results. Calvo~et al.~\cite{Calvo} collected 14 684 ads from six parties during Spain's 2019 general election. They manually coded 1 743 ads according to 9 topics of interest to understand how much money different parties spend on promoting different topics. 
Dobber~et~al.~\cite{dobber} analyzed the electoral promises Dutch political parties were making during the 2019 European elections. The authors collected and labeled 362 ads according to the CAP codebook. 
Their analysis showed that political campaigns promoted electoral promises only to small groups of people and concluded that this is problematic from a democratic accountability perspective. These sorts of questions are what motivated us to propose automated methods that enable robust and large-scale analyses.  
Flower~et~al.~\cite{Fowler} showed, by comparing Facebook posts from 7 056 candidates and T.V. ads from 1 274 candidates in the 2018 U.S. mid-term election, that Facebook posts are used for a more diverse range of goals--such as fundraising than are TV ads. 
Finally, Gitomer~et~al.~\cite{gitomer2021geographic} analyzed location targeting of political ads in Germany.  In our work, we dig deeper and analyze if different policy categories receive more/less attention across different regions in France. 

Manual labeling of large amounts of ads is time-consuming and costly, a few recent works have proposed methods to \emph{automatically label ads}.  Baviera~et~al. \cite{Baviera} used the Key-phrase Digger algorithm~\cite{Moretti2015DiggingIT} to detect the main terms in the texts of the 14 684 Facebook ads. They found that the main aim of the ads was to mobilize voters. This work is orthogonal to ours as it provides a less comprehensive but more focused perspective on topics discussed. 
Regarding video content, Baskota~et~al.~\cite{imabalnce_data} proposed methods to classify the tone in political videos and 
Banerjee~et~al.~\cite{video_classification} proposed methods to differentiate political campaign ads from other online video ads. 

Finally, a few works have audited the transparency mechanisms provided by ad platforms. 
Marcio~et~al.~\cite{Marcio} showed that around half of the political ads send in Brazil during the 2018 presidential election were missing from the Meta's Ad Library. Le Pochat~et~al.~\cite{Victor} audited Meta's political ads reinforcements and showed that Meta's algorithms can detect more than 40\% of unlabeled ads in less than 24 hours. Furthermore, Edelson~et~al.~\cite{Laura} conduct a security audit for the U.S. Ad Library and discovered that 54.6\% of pages never provide a disclosure string. 
Overall, these articles showed that even if Meta Ad Library is an excellent tool that brings more transparency, it has big design and implementation flaws. 

\vspace{1mm}
\textit{Works on online political content analysis.}
More broadly, a lot of research in both political and communication sciences has analyzed online political content on social media.  Because of space constraints we only discuss works related to policy analysis. 

%

Rusell~et~al.~\cite{Russell} examined what women in the U.S. Congress discuss on Twitter. She collected 113 112 tweets from verified senator's accounts and trained students to \textit{manually label} them according to 20 major topics from U.S. Policy Agenda Project coding scheme. The results showed that congresswomen post on Twitter about diverse topics and do not focus only on women-related issues.

The biggest problem when building \textit{automated methods to label political text} is the lack of labeled data. 
Terechshenko~et~al.~\cite{terechshenko} propose to use transfer learning and showed that RoBERTa achieved the highest accuracy score of 61\% when trained on  the CAP bills dataset and tested on the  CAP New York Times headlines.
In this paper we showed that transfer learning from CAP bills to ads results in very low accuracy. 
Hemphill~et~al.~~\cite{Hemphill} investigated policy attention among different U.S. congress members on Twitter. They manually labeled 59k tweets according to the CAP scheme. Using logistic regression with bag of words they achieved a 0.79 F1. They found that the proportion of congress members' tweets about policy issues stayed stable. The paper does not provide any details on the annotation process and does not show the accuracy across different policy categories. In fact, some of the results could be invalidated if the recall differs for different policy categories as we show it happens in the context of political ads. 
Finally, Jackson~et~al.~\cite{jackson} proposed to use a lexicon-based approach to built a list of language cues for nine political topics to deal with the lack of training data. 
The authors evaluated the method over 500 labeled texts and they achieved an accuracy of over 85\% for eight out of nine categories.  Gupta~et~al.~\cite{gupta2020polibert} used a supervised approach for classifying ads into different categories such as advocacy, attack, image, and issue; but without investigating the precise issue discussed. The authors manually labeled 5 231 Tweets and 4 434 Facebook posts which they used to build a BERT classifier that achieves an accuracy of 83\%. 
Overall, there have been several related works on analyzing political content, however, none of them provides the solid foundations we provide for analyzing policy-related ads that goes from having the right codebooks, investigating difficulties in annotation and understanding which language models configurations are most suitable for supporting such nuanced classification. 
Our paper is also the first to analyze policy attention in political ads at large-scale and show imbalances across demographic groups.

%% file: conclusion.tex
\section{Conclusion}

Given the large volumes of data available, there is an increasing need for automatic methods to investigate paid political speech. 
This paper explores automated methods to label political ads according to 14 policy categories. Understanding policy attention is important for analyzing  democratic processes. 
Our models are able to achieve over 0.75 F1 scores for policy categories such as \textit{environment} and \textit{cultural policy} and F1 scores between 0.5 and 0.7 for policy categories such as \textit{energy} and {health}. Overall, the categories with high accuracy correspond to those with a higher agreement between Prolific annotators. The main culprit for disagreement are the ads who's messages relates to multiple policy categories. 
Our methods could be used in conjunction with methods to detect sentiment and tone to identify deceiving political ads that exploit vulnerable groups of people through targeting. 

Finally, we build one of the few models in the literature to analyze political content in French. Using this model we analyzed the online ads posted during the 2022 French Presidential Election. We observe significant imbalances in the policies discussed in ads that target different demographic groups. Such imbalances could affect voters deliberation and, hence, need to be taken into account when designing political ad targeting technologies.   


%% file: Appendix.tex
\appendix
\section{Appendix}
\label{appendix}
\newpage
\paragraph{Data available in the Meta Ad Library on ads}
For each ad, Meta's ad library provides the:
\begin{itemize}
    \item \textit{creation time} of an ad.
    \item \textit{creative body}--a text of an ad.
    \item \textit{bylines}--the information about who paid for an ad, that advertisers are required to provide.
    \item \textit{demographic distribution} -- information about the age and gender of people reached by an ad.
    \item \textit{region distribution}--distribution of people reached by an ad over regions in France.
    \item \textit{impressions}--a field that shows the number of times the ad created an impression.
    \item \textit{language}--the list of languages of the texts of the ad.
    \item \textit{currency}, that was used to pay for an ad.
    \item \textit{spend} -- the amount of money spent running the ad as specified in currency.
\end{itemize}

\begin{table}[t]
\centering
\caption{Accuracy when the gold labels are considered ground truth, and the Prolific labels are considered predictions.}
\footnotesize{
\begin{tabular}{lrrrr}
    \toprule
  &Prec. & Rec. & F-1 & Support  \\
  \midrule
International affairs & 0.61 & 0.74 & 0.67&50 \\
Energy & 0.77 &0.81&0.79&68 \\
Government operations & 0.74 & 0.58&0.65&85\\
Cultural policy & 0.84& 0.65&0.73&80\\
Social policy & 0.39& 0.56&0.46&48\\
Health&0.68&0.77&0.72&56\\
Human rights & 0.45&0.78&0.57&49\\
Environment & 0.62& 0.78&0.69&78\\
Economy&0.40&0.45&0.42&53\\
\bottomrule
\end{tabular}}
\label{tab:test-classificationReportGold}
\end{table}

\begin{figure*}[!]
    \centering
        \caption{Policy category heat map for Gold and Prolific labels.}
    \includegraphics[width=12cm]{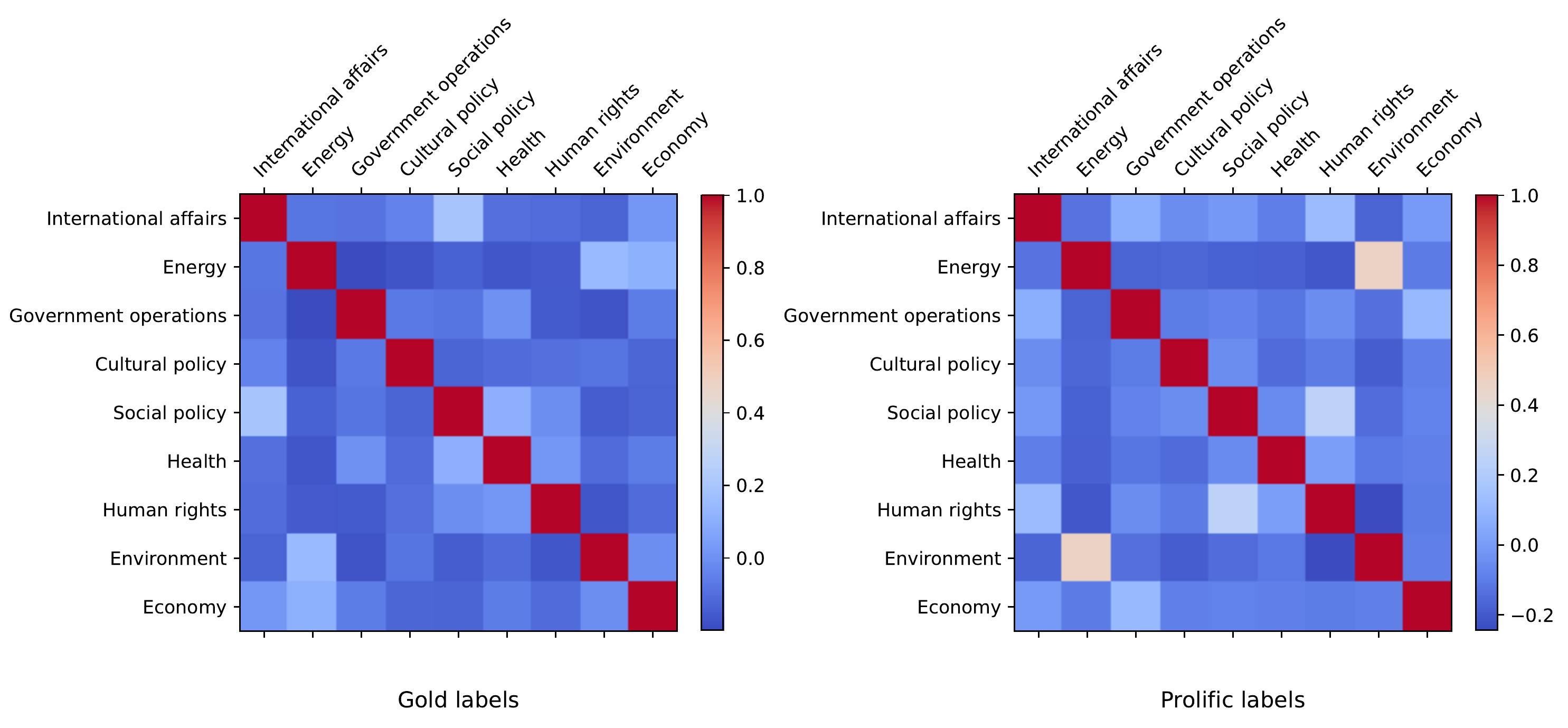}
    \label{fig:heat_map_before_gold}
\end{figure*}

\begin{table}[t]
\centering
\caption{Accuracy across policy categories using FlauBERTtuned over the \testf. \textbf{Different thresholds per category.}}

\footnotesize{
\begin{tabular}{lrrrr}
    \toprule
  &Prec. & Rec. & F-1 & Support  \\
  \midrule
International affairs & 0.93 & 0.26 & 0.41&102 \\
Energy & 0.96 &0.45&0.61&100 \\
Immigration&0.57&0.27&0.36&30\\
Law and crime&0.75&0.26&0.38&35\\
Government operations & 0.73 & 0.21&0.33&105\\
Cultural policy & 0.78& 0.75&0.76&110\\
Social policy & 0.76& 0.12&0.21&104\\
Urban and territorial policies&0.89&0.26&0.40&31\\
Health&0.84&0.60&0.70&101\\
Labor and employment & 1&0.17&0.29&36\\
Human rights & 0.79&0.20&0.32&134\\
Education&0.75&0.29&0.42&31\\
Environment & 0.79& 0.71&0.75&150\\
Economy&0.89&0.16&0.26&103\\
\midrule
micro avg&0.81&0.38&0.51&1172\\
samples avg&0.52&0.43&0.46&1172\\
\bottomrule
\end{tabular}
}
\label{tab:test-14cat-tho}
\end{table}

\begin{table*}[!]
\caption{Distribution of ad impressions across regions and policy categories. $^*$ represents the region distribution of impressions for all ads in French that have at least one predicted policy.}
\resizebox{18cm}{!}{
\begin{tabular}{l|c|c|c|c|c|c|c|c|c|c|c|c|c}
                      \toprule
                             & \multicolumn{13}{c}{\textbf{Regions}}  \\
\hline
& \multicolumn{1}{m{2cm}|}{Auvergne-Rhône-Alpes} & \multicolumn{1}{m{2cm}|}{Bourgogne Franche-Comté} & Bretagne & \multicolumn{1}{m{2cm}|}{Centre-Val de Loire} & Corse  & Grand Est & \multicolumn{1}{m{2cm}|}{Haut De France} & Narmandie &\multicolumn{1}{m{1.5cm}|}{Nouvelle-Aquitaine} & Occitanie & \multicolumn{1}{m{1.8cm}|}{Pays De La Loire} & \multicolumn{1}{m{2cm}|}{Provence Alpes Côte D'Azur} & Ile-De-France \\
\midrule
Population$^*$                   & 11.7\%               & 4.25\%                  & 4.96\%   & 3.78\%              & 0.76\% & 8.22\%    & 9.60\%         & 5.12\%    & 9.69\%             & 10.07\%   & 5.39\%           & 8.89\%                     & 17.55\%       \\
International affairs        & 12.25\%              & 3.76\%                  & 4.94\%   & 3.66\%              & 0.80\% & 8.32\%    & 8.43\%         & 4.92\%    & 9.69\%             & 10.07\%   & 4.90\%           & 9.46\%                     & 18.79\%       \\
Energy                       & 6.31\%               & 6.23\%                  & 1.41\%   & 6.50\%              & 0.10\% & 6.80\%    & 15.85\%        & 7.54\%    & 11.73\%            & 7.33\%    & 6.69\%           & 6.71\%                     & 16.81\%       \\
Immigration                  & 12.30\%              & 3.98\%                  & 5.09\%   & 3.56\%              & 0.81\% & 7.98\%    & 8.23\%         & 4.78\%    & 9.58\%             & 10.24\%   & 4.79\%           & 9.75\%                     & 18.91\%       \\
Justice and ciminalty        & 5.50\%               & 1.92\%                  & 2.96\%   & 1.74\%              & 0.49\% & 3.75\%    & 3.69\%         & 2.51\%    & 5.17\%             & 5.37\%    & 2.59\%           & 4.74\%                     & \cellcolor{bittersweet}{59.57\%}       \\
Gouvernement operations      & 11.76\%              & 4.03\%                  & 5.14\%   & 3.74\%              & 0.79\% & 8.47\%    & 9.89\%         & 5.59\%    & 9.89\%             & 9.43\%    & 5.50\%           & 8.20\%                     & 17.57\%       \\
Cultural policy              & 11.04\%              & 3.60\%                  & 5.17\%   & 3.46\%              & 0.78\% & 7.23\%    & 7.85\%         & 4.99\%    & 9.38\%             & 10.98\%   & 5.87\%           & 9.19\%                     & 20.47\%       \\
Social Policy                & 11.46\%              & 4.27\%                  & 5.05\%   & 3.88\%              & 0.78\% & 8.77\%    & 10.53\%        & 5.46\%    & 9.76\%             & 9.81\%    & 5.22\%           & 8.95\%                     & 16.05\%       \\
Education                    & 10.72\%              & 3.79\%                  & 4.86\%   & 3.74\%              & 0.63\% & 7.82\%    & 8.90\%         & 4.95\%    & 9.09\%             & 12.90\%   & 5.25\%           & 7.80\%                     & 19.55\%       \\
Environment                  & 10.66\%              & 5.39\%                  & 4.59\%   & 4.09\%              & 0.64\% & 8.33\%    & 10.43\%        & 5.34\%    & 10.50\%            & 10.06\%   & 5.57\%           & 8.63\%                     & 15.76\%       \\
Health                       & 11.17\%              & 4.09\%                  & 4.68\%   & 3.88\%              & 0.85\% & 8.63\%    & 10.51\%        & 5.34\%    & 9.39\%             & 9.46\%    & 4.93\%           & 9.81\%                     & 17.27\%       \\
Economy                      & 10.55\%              & 4.30\%                  & 9.31\%   & 3.50\%              & 1.29\% & 8.39\%    & 10.38\%        & 5.00\%    & 8.66\%             & 8.75\%    & 4.54\%           & 8.82\%                     & 16.51\%       \\
Human rights                & 13.70\%              & 3.87\%                  & 5.29\%   & 3.66\%              & 0.65\% & 7.90\%    & 9.42\%         & 5.05\%    & 9.59\%             & 9.92\%    & 5.51\%           & 8.43\%                     & 17.01\%       \\
Work and employment          & 10.95\%              & 4.09\%                  & 4.88\%   & 4.41\%              & 0.63\% & 8.66\%    & 11.92\%        & 5.61\%    & 9.12\%             & 8.75\%    & 5.33\%           & 9.59\%                     & 16.07\%       \\
Urban, territorial policy & \cellcolor{bittersweet}{27.08\%}              & 4.87\%                  & 2.42\%   & 2.25\%              & 0.46\% & 14.26\%   & 6.74\%         & 3.13\%    & 6.75\%             & 6.67\%    & 2.60\%           & 11.18\%                    & 11.60\%      \\
\bottomrule
\end{tabular}}
\label{tab:region_policy}
\end{table*}

\begin{table*}[]
\caption{Examples of ads that caused disagreement between Prolific workers and experts.}
\tiny{
\begin{tabular}{|p{0.15\textwidth}|p{0.35\textwidth}|p{0.35\textwidth}|}
\hline
Category & False positive   & False negative  \\
\hline

International affairs &Do you like Portugal? You are going to love 2022. More than 200 events to celebrate Franco-Portuguese friendship: music, cinema, visual arts, theatre, cinema, literature, gastronomy. discover contemporary Portugal! Support the France-Portugal 2022 Season and don't miss any event by subscribing to the page! 

\textbf{experts' label}: cultural policy; \textbf{prolific workers' label}: international affairs.

& Afghans, Syrians, Sudanese... More than 26 million refugees have fled violence and persecution around the world.26 million, but as many unique stories, life paths and future projects. For 50 years, France Terre d'Asile works to defend the right to asylum and accompanies those who seek protection in France. We need you to continue!

\textbf{experts' labels}: international affairs, human rights; \textbf{prolific workers' labels}: social policy, human rights.\\

\hline
Energy & It's official, today we say goodbye to winter and hello to spring! What if we take advantage of this new season to take care of nature?     

\textbf{experts' labels}: environment; \textbf{prolific workers' labels}: energy, environment.
& The gas we consume today allows Putin to finance his war. Tomorrow, we will have to manage to do without it. But right now we can bring down the temperature… and the bill. For Ukraine, I'm wearing my \#PatrioticSweater and turning down the heat.  

\textbf{experts' labels}: energy, international affairs; \textbf{prolific workers' labels}: economy.\\
                                      \hline  
Economy              & With the crisis in Ukraine, some states wished to join the EU, in particular to prevent the conflict from being exported to their borders. Concretely, how can a country join the EU?           

\textbf{experts' labels}: international affairs; \textbf{prolific workers' labels}: international affairs, economy.

& Banks must stop massively financing, without our knowledge, fossil fuels and polluting industries that aggravate global warming, the decline of biodiversity and therefore the living conditions of people. Sign the manifesto for a sustainable and transparent bank and regain power over your money.

\textbf{experts' labels}: energy, environment, economy; \textbf{prolific workers' labels}: environment.

\\

        \hline      
\end{tabular}}
\label{tab:exampl_disagr}
\end{table*}